\documentclass[journal]{IEEEtran}
\usepackage{amsmath,graphicx}

\usepackage{booktabs,array}
\usepackage{amsmath} % assumes amsmath package installed
\usepackage{amssymb}  % assumes amsmath package installed
\usepackage{array}
\usepackage{amsthm}
\usepackage{rotating}
\usepackage{rotfloat}
\usepackage{enumitem}
\usepackage[export]{adjustbox}
\usepackage{arydshln}
\usepackage{float}
\usepackage{tabularx}
\usepackage{tabularx}
\usepackage{multirow}
\usepackage[numbers,sort&compress]{natbib}
\usepackage{amsmath}
\usepackage{enumitem}
\usepackage{bm}
\usepackage{textcomp}
\usepackage{url}

\usepackage{booktabs}% http://ctan.org/pkg/booktabs
\newcommand{\tabitem}{~~\llap{\textbullet}~~}
\usepackage{tabularx}
\usepackage{amsmath}
\ifCLASSINFOpdf
\else
\fi
\begin{document}
\title{Iterative Methods for Sparse Signal Reconstruction from Level Crossings}

\author{Mahdi~Boloursaz~Mashhadi,~\IEEEmembership{Student Member,~IEEE,}~and~Farokh~Marvasti,~\IEEEmembership{Senior~Member,~IEEE}%  
\thanks{The authors are with the Advanced Communications Research Institute (ACRI), EE Department, Sharif University of Technology (SUT), Tehran, Iran, e-mail: boloursaz@ee.sharif.edu}}


\maketitle

\begin{abstract}
This letter considers the problem of sparse signal reconstruction from the timing of its Level Crossings (LC)s. We formulate the sparse Zero Crossing (ZC) reconstruction problem in terms of a single 1-bit Compressive Sensing (CS) model. We also extend the Smoothed L0 (SL0) sparse reconstruction algorithm to the 1-bit CS framework and propose the Binary SL0 (BSL0) algorithm for iterative reconstruction of the sparse signal from ZCs in cases where the number of sparse coefficients is not known to the reconstruction algorithm a priori. Similar to the ZC case, we propose a system of simultaneously constrained signed-CS problems to reconstruct a sparse signal from its Level Crossings (LC)s and modify both the Binary Iterative Hard Thresholding (BIHT) and BSL0 algorithms to solve this problem. Simulation results demonstrate superior performance of the proposed LC reconstruction techniques in comparison with the literature. 
\end{abstract}


\IEEEpeerreviewmaketitle

\section{Introduction}
\IEEEPARstart{U}{niform} sampling is a popular strategy in the conventional Analog to Digital (A/D) converters. However, an alternative technique could be Level Crossing (LC) sampling \cite{MarkTodd, Sayiner, AllierSicard, Tsividis2} which samples the input analog signal whenever its amplitude crosses any of a predefined set of reference levels. LC based A/Ds represent each LC by encoding its quantized time instance along with an additional bit that represents the value of the level crossed at that time instant \cite{AllierSicard}.  

LC sampling generates signal-dependent non-uniform samples and benefits from certain appealing properties in comparison with the conventional uniform sampling technique. It reduces the number of samples by automatically adapting the sampling density to the local spectral properties of the signal \cite{Sekhar, Greitans}. Furthermore, LC based A/Ds can be implemented asynchronously and without a global clock. This in turn leads to reduced power consumption, heating and electromagnetic interference \cite{Akopyan,Kinniment}.  

A seminal work by Logan \cite{Logan} showed that signals with octave-band Fourier spectra can be uniquely reconstructed from their zero crossings up to a scale factor. This is a sufficient but not necessary condition for LC signal reconstruction. Previous works on LC signal reconstruction have mostly considered low \cite{Chegini,Lazar} or band pass \cite{Logan} signal assumption and there are few prior works that utilize sparsity \cite{Sharma,Boufounos}. Boufounos et. al. \cite{Boufounos} formulates the zero crossing reconstruction problem as minimization of a sparsity inducing cost function on the unit sphere and Sharma et. al. \cite{Sharma} uses the Basis Pursuit (BP) and Orthogonal Matching Pursuit (OMP) \cite{OMP} techniques to reconstruct the signal from LC samples. Both \cite{Sharma, Boufounos} formulate the LC reconstruction problem in terms of a conventional Compressive Sensing (CS) \cite{Candes} reconstruction model.

\textbf{\textit{Contributions}}: In this work, we utilize the emergent theory of 1-bit CS \cite{1bitBoufounos, 1bitJacques} to formulate the LC problem. We show how the LC problem can be addressed by a system of simultaneously constrained signed-CS problems and modify the Binary Iterative Hard Thresholding (BIHT) and Binary Smoothed L0 (BSL0) algorithms to solve this problem. 

For further reproduction of the results reported in this paper, MATLAB files are provided online at ee.sharif.edu/$\sim$boloursaz.

The rest of this paper is organized as follows. In section \ref{sec:formulation} we formulate the LC problem in terms of 1-bit CS models. Section \ref{sec:PropAlgs} presents the proposed BSL0 and the modified BIHT and BSL0 algorithms. Section \ref{sec:sim} provides the simulation results and finally section \ref{sec:con} concludes the paper.

\section{Problem Formulation}
\label{sec:formulation}
In this section, we formulate the problem of sparse signal reconstruction from level crossings and address the similarities and differences between this problem and a typical 1-bit CS problem.
\subsection{Zero Crossing (ZC) Reconstruction}\label{subsec:ZerocrossMod}
Suppose $x(t)=\sum_{n=0}^{N} a_n \cos(n \omega_0 t)$, for $t \in [0,d]$. Also define the spectral support as $\mathcal{S}=\{n|a_n \ne 0\}$. Now, the sparse signal assumption imposes that $K=|\mathcal{S}|<<N$. Also denote by $x[m]=x(mT), m=0, 1, ..., M-1$ the uniform samples taken from $x(t)$ at rate $1/T<<N\omega_0/\pi$ significantly below Nyquist in which $(M-1)T=d$. It is obvious that a ZC-based A/D can extract $y(t)=sign(x(t))$ from the ZC time instances and the initial sign of $x(t)$. Hence, we have $y[m]=sign(x[m])$. Now in vector notation we can write (\ref{form1})
\begin{align}\label{form1}
	\textbf{y}=sign(\textbf{x})=sign(\Phi \textbf{a})
\end{align} 

in which the vector $\textbf{x}_{M \times 1}=[x[0] \ x[1] \ ... \ x[M-1]]^T$ contains the uniform samples and $\textbf{y}_{M \times 1}=[y[0] \ y[1] \ ... \ y[M-1]]^T$ contains the corresponding sign values. The vector $\textbf{a}_{(N+1) \times 1}=[a_0 \ a_1 \ ... \ a_N]^T$ contains the sparse coefficients and

\begin{align}
\Phi_{M \times (N+1)}=\left( {\begin{array}{*{20}{c}}
	1\\
	\vdots\\
	1
	\end{array}\begin{array}{*{20}{c}}
	{\cos(2\omega_0 T)}& \ldots &{\cos(N\omega_0 T) }\\
	\vdots & \ddots & \vdots \\
	{\cos(2\omega_0MT)}& \cdots &{\cos(N\omega_0 MT) }
	\end{array}} \right)\nonumber
\end{align}

Note that in (\ref{form1}), we need to estimate the sparse coefficient vector $\textbf{a}$ from the sign measurements $\textbf{y}$. Of course, reconstruction is only possible up to a scale factor. Hence, we need to add the norm constraint  $||\textbf{a}||_2=1$ which yields a typical 1-bit CS problem (\ref{onebitprob}) that can be solved by the Matching Sign Pursuit (MSP) \cite{MSP}, Binary Iterative Hard Thresholding (BIHT) \cite{1bitJacques}, 1-bit Bayesian Compressive Sensing \cite{1bitBCS} or any other 1-bit CS reconstruction algorithm. Once $\textbf{a}$ is estimated, the sparse analog signal $x(t)$ is estimated at infinite accuracy.  

\begin{equation}\label{onebitprob}
	\begin{array}{rrclcl}
		\displaystyle \min_{\textbf{a}} & \multicolumn{3}{l}{\|\textbf{a}\|_0} \\
		\textrm{s.t.} &&&\\
		& \textbf{y} & = & sign(\Phi \textbf{a}) \\
		& \|\textbf{a}\|_2 & = & 1 \\
	\end{array}
\end{equation}

\subsection{Level Crossing (LC) Reconstruction}\label{subsec:LevelcrossMod}

Now consider the multi-level scenario in which the temporal instances of the signal crossings with a predefined set of reference levels is encoded and transmitted to the receiver. Let’s denote the set of levels by $\mathcal{L}=\{l_{-L/2}, ..., l_0, ..., l_{L/2}\}$. Similar to the single-level case (zero crossings), the temporal instances of the crossings provides the following signals (\ref{form2})
\begin{align}\label{form2}
	y_{L/2}(t)&=sign(x(t)-l_{L/2})\nonumber\\
	&\vdots\nonumber\\
	y_0(t)&=sign(x(t)-l_0)\nonumber\\
	&\vdots\nonumber\\
	y_{-L/2}(t)&=sign(x(t)-l_{-L/2})
\end{align}

Similarly, let’s denote by $x[m]=x(mT),   m=0,1,…,M-1$ the uniform samples taken from $x(t)$ at rate $1/T<<N\omega_0/\pi$ significantly below Nyquist. Now in vector notation we can write (\ref{form3})
\begin{align}\label{form3}
	\textbf{y}_{L/2}&=sign(\textbf{x}-l_{L/2})=sign(\Phi \textbf{a}-l_{L/2})\nonumber\\
	&\vdots\nonumber\\
	\textbf{y}_0&=sign(\textbf{x}-l_0)=sign(\Phi \textbf{a}-l_0)\nonumber\\
	&\vdots\nonumber\\
	\textbf{y}_{-L/2}&=sign(\textbf{x}-l_{-L/2})=sign(\Phi \textbf{a}-l_{-L/2})
\end{align}

in which the vectors $\textbf{x}$ and $\textbf{a}$ and the matrix $\Phi$ are the same as defined in subsection \ref{subsec:ZerocrossMod} and the vectors $\textbf{y}_{-L/2}, ...,\textbf{y}_0, ..., \textbf{y}_{L/2}$ contain the corresponding sign values. Now in order to solve the above system of signed-CS problems simultaneously, we define the vector $\textbf{y}'$ as (\ref{form4})
\begin{equation}\label{form4}
	\textbf{y}'=\left( \begin{array}{l}
		\textbf{y}_{L/2}\\
		\vdots \\
		\textbf{y}_0\\
		\vdots \\
		\textbf{y}_{-L/2}
	\end{array} \right)=sign(\Phi'\textbf{a}')
\end{equation}
in which $\Phi'_{(M(L+1)) \times (N+L+2)}$ is made by concatenation of the $\Phi$ matrices and the level vectors  according to (\ref{form5})
\begin{equation}\label{form5}
	\Phi'=\left(\begin{array}{c c c c c c}
		\Phi & (l_{L/2})_{M \times 1} & \ldots & (0)_{M \times 1} & \ldots & (0)_{M \times 1}\\
		&   & \vdots &   & \vdots &  \\
		\Phi & (0)_{M \times 1} & \ldots & (l_0)_{M \times 1} & \ldots & (0)_{M \times 1}\\
		&   & \vdots &   & \vdots &  \\
		\Phi & (0)_{M \times 1} & \ldots & (0)_{M \times 1} & \ldots & (l_{-L/2})_{M \times 1}\\
	\end{array}\right)
\end{equation}

In (\ref{form5}), each level vector $(l_i)_{M \times 1}$ is a column vector with all entries equal to the level $l_i$. Finally, $(0)_{M \times 1}$ denotes all zero column vectors and $\textbf{a}'_{(N+L+2) \times 1}=\left(\begin{array}{c}
\textbf{a}\\(-1)_{(L+1) \times 1}
\end{array}\right)$. 
Hence, to estimate the sparse vector of coefficients $\textbf{a}$, we need to solve the constrained signed-CS problem (\ref{LevelCSprob}). Note that wherever we replace the term "1-bit CS"  with "signed-CS" throughout this paper (e.g. in referring to (\ref{LevelCSprob})), we are emphasizing the difference between that problem and a typical 1-bit CS problem regarding the scaling ambiguity. For example, problem (\ref{LevelCSprob}) is well-posed and does not need the additional norm constraint.

\begin{equation}\label{LevelCSprob}
	\begin{array}{rrclcl}
		\displaystyle \min_{\textbf{a}'} & \multicolumn{3}{l}{\|\textbf{a}'\|_0} \\
		\textrm{s.t.} &&&\\
		& \textbf{y}' & = & sign(\Phi' \textbf{a}') \\
		& \textbf{a}'_{N+2:N+L+2} & = & (-1)_{(L+1) \times 1} \\
	\end{array}
\end{equation}
In section (\ref{sec:PropAlgs}), we propose efficient algorithms to solve (\ref{LevelCSprob}).

\section{The Proposed Algorithms}
\label{sec:PropAlgs}

In this section, we present our proposed algorithms. In subsection \ref{subsec:BSL0} we present the Binary Smoothed L0 (BSL0) algorithm proposed for solving (\ref{onebitprob}) in case where the sparsity number $K$ is not known for reconstruction. Subsequently in subsection \ref{subsec:SparseLCProb}, we present our proposed algorithms for solving the sparse LC problem (\ref{LevelCSprob}). 

\subsection{The Binary Smoothed L0 (BSL0) Algorithm}\label{subsec:BSL0}
Some previously proposed 1-bit CS reconstruction algorithms (e.g. BIHT) need prior knowledge of the sparsity number $K$ for reconstruction. However, $K$ is usually not known to the reconstruction algorithm in the real-world scenario considered in this paper. To cope with this problem, we propose the Binary Smoothed L0 (BSL0) algorithm. Note that although the simulation results for BSL0 are provided for the ZC/LC scenario in this paper, the algorithm is also applicable to the general scenario of 1-bit CS. 

The basic SL0 algorithm was proposed in \cite{SL01, SL02}, for finding sparse solutions to 	under-determined systems of linear equations. The main idea of SL0 is to apply the Graduated Non-Convexity (GNC) technique \cite{GNC} and approximate the discontinuous $l^0$ norm by a sequence of continuous functions to enable using continuous minimization techniques. In this work, we apply the same idea to find the solution to the 1-bit CS problem as stated in (\ref{onebitprob}). To this end, we solve the following problem iteratively (\ref{BSL0Form}) 
\begin{equation}\label{BSL0Form}
	\begin{array}{rrclcl}
		\displaystyle \min_{\textbf{a}} & \multicolumn{3}{l}{C_{\sigma,\lambda, \theta}(\textbf{a})=F_\sigma(\textbf{a})+ \lambda J(\textbf{a})+ \theta (\|\textbf{a}\|_2^2-1)^2}\\
	\end{array}
\end{equation}
in which $J(\textbf{a})=\|[Y(\Phi\textbf{a})]_-\|_1$, $Y=diag(\textbf{y})$ and $[.]_-$ denotes the negative function, i.e., $([\textbf{a}]_-)_i=[a_i]_-$ with $[a_i]_-=a_i$ if $a_i<0$ and $0$ else. Also, we have $\lim\limits_{\sigma \to 0^+} F_{\sigma}(\textbf{a})=\|\textbf{a}\|_0$.

Note that the first term of the cost function ($F_\sigma(\textbf{a})$) enforces sparsity, the second term ($J(\textbf{a})$) enforces consistency of the solution to the set of sign measurements and ($(\|\textbf{a}\|_2^2-1)^2$) enforces the final solution to be located on the unit sphere to avoid scaling ambiguity. The idea is to decrease $\sigma$ along the iterations to better approximate the $l^0$-norm while increasing $\lambda$ and $\theta$ to enforce the sign and norm constraints.

The proposed BSL0 algorithm takes a dual loop approach to solve (\ref{BSL0Form}). Similar to the basic SL0 \cite{SL01}, the inner loop is a Gradient Descent algorithm that is applied on the sequence of cost functions $C_{\sigma_0,\lambda_j,\theta_j}(\textbf{a}), C_{\sigma_1,\lambda_j,\theta_j}(\textbf{a}), ..., C_{\sigma_k,\lambda_j,\theta_j}(\textbf{a}),$ where $\sigma_i=\alpha\sigma_{i-1}, 0<\alpha<1$. In each iteration of the outer loop, the $\lambda, \theta$ parameters are increased by $\lambda_j=\beta\lambda_{j-1}$ and $\theta_j=\delta\theta_{j-1}$ where $1<\beta,\delta$. 

As stated in \cite{SL01}, there exists several different choices for the $l^0$-norm approximation function ($F_{\sigma}(\textbf{a})$) and in this research, we assume $F_{\sigma}(\textbf{a})=\sum_{m=0}^{N} (1-exp(-a_m^2/\sigma^2))$. Hence, considering a set of fixed parameters ($\sigma, \lambda, \theta$) for the inner gradient descent algorithm we have (\ref{SubGrad})

\begin{align}\label{SubGrad}
	\nabla C_{\sigma, \lambda, \theta}(\textbf{a})&=\frac{2}{\sigma^2}\left(\begin{array}{c c c c}
		e^{-a_0^2/\sigma^2}&0&\ldots&0\\0&e^{-a_1^2/\sigma^2}&\ldots&0\\\vdots&\vdots&\vdots&\vdots\\0&\ldots&0&e^{-a_N^2/\sigma^2}
	\end{array}\right)\textbf{a}\nonumber\\
	&+\frac{\lambda}{2}\Phi^T(sign(\Phi\textbf{a})-\textbf{y})+\theta ((||\textbf{a}||_2^2-1))\textbf{a}
\end{align}
Precisely speaking, (\ref{SubGrad}) is in fact a sub-gradient of the cost function because the second term ($\frac{\lambda}{2}\Phi^T(sign(\Phi\textbf{a})-\textbf{y})$) is a sub-gradient of $ \lambda J(\textbf{a})$ as proved in \cite{1bitJacques}. \textbf{Algorithm 1} gives the formal presentation of the proposed BSL0 algorithm. 

\subsection{The Sparse LC Problem}\label{subsec:SparseLCProb}
In this subsection, we modify both the BIHT \cite{1bitJacques} and the BSL0 algorithms to solve the sparse LC problem (\ref{LevelCSprob}). Note that the only difference between the sparse ZC model formulated in (\ref{onebitprob}) and the sparse LC model (\ref{LevelCSprob}) is the constraint on the sparse coefficient vector $\textbf{a}'$. Also note that $\mathcal{C}=\{\textbf{a}' \in R^{N+L+2}\,|\, \textbf{a}'_{N+2:N+L+2}=(-1)_{(L+1) \times 1}\}$ the set of all real vectors with last $L+1$ entries equal to $-1$ is convex. Hence, to enforce this constraint, we can simply project the solution onto $\mathcal{C}$ at each iteration. As $\mathcal{C}$ is convex, this projection will not hamper convergence of the overall iterative algorithm. 

For the modified BSL0 we solve (\ref{MBSL0})

\begin{equation}\label{MBSL0}
	\begin{array}{rrclcl}
		\displaystyle \min_{\textbf{a}'} & \multicolumn{3}{l}{F_\sigma(\textbf{a}')+ \lambda J(\textbf{a}')} \\
		\textrm{s.t.} &&&\\
		& \textbf{a}'_{N+2:N+L+2} & = & (-1)_{(L+1) \times 1} \\
	\end{array}
\end{equation}

\begin{table}[H]
	\centering
	\begin{tabular}{|l|l|}
		\hline
		%		\multicolumn{1}{|c|}{ }\\ 
		\multicolumn{1}{|c|}{\textbf{Algorithm 1} Binary Smoothed L0 (BSL0)}\\
		%		\multicolumn{1}{|c|}{ }\\ 
		\hline \\	
		\textbf{Inputs}: \\
		\quad\tabitem 	$\Phi_{M\times(N+1)}$: The sampling matrix\\
		\quad\tabitem $\textbf{y}_{M\times1}$: The vector of sign measurements\\
		\quad\tabitem \textit{$\epsilon$}: The stopping criteria\\
		\quad\tabitem $\mathbf{(\sigma_0,\lambda_0,\theta_0)}$: The initial algorithm parameters\\
		\quad\tabitem $\mathbf{(\alpha,\beta,\delta)}$: The parameter increase/decrease factors\\ 
		\quad\tabitem $IterMax$: The maximum number of iterations\\ 
		\quad\tabitem $\sigma_{min}$: The minimum $\sigma$ parameter allowed\\ 
		\quad\tabitem $\mu$: The step-size to the Gradient Descent (GD)\\   
		
		\textbf{Output}: \\
		\quad\tabitem 
		$\hat{\textbf{a}}^{(k)}$: The estimated vector of sparse coefficients\\
		\textbf{Algorithm}: \\
		\quad\tabitem Initialization \textit {$\hat{\textbf{a}}^{(1)}=[0]_{(N+1)\times 1},\hat{\textbf{a}}^{(0)}=[-100]_{(N+1) \times 1}$}\\
		\quad \quad \textit{$k=1, i=j=0$}\\
		\quad\tabitem \textbf{While} $(||\hat{\textbf{a}}^{(k)}-\hat{\textbf{a}}^{(k-1)}||>\epsilon)$ and ($k<IterMax$) \\
		
		\quad\quad\tabitem \textbf{While} $(\sigma_i>\sigma_{min})$  \\		
		\quad \quad  \quad  -~Calculate the gradient vector $\nabla C_{\sigma_i, \lambda_j,\theta_j}(\hat{\textbf{a}}^{(k)})$ (\ref{SubGrad}) \\
		\quad \quad  \quad  -~Perform the gradient descent (GD) step as:\\

		\begin{tabular}{p{8cm}}
			
			%	\tableequation 
			{\quad \quad\quad \quad  \quad $\hat{\textbf{a}}^{(k+1)}=\hat{\textbf{a}}^{(k)}-\mu \nabla C_{\sigma_i, \lambda_j,\theta_j}(\hat{\textbf{a}}^{(k)})$}\\

		\end{tabular}\\
		\quad \quad  \quad   -~$\sigma_{i+1}=\alpha\sigma_{i}$ \\
		\quad \quad  \quad   -~$i=i+1$ \\
		\quad \quad  \quad   -~$k=k+1$ \\
		\quad\quad\tabitem End \textbf{While}\\
		\quad\quad\tabitem $\lambda_j=\beta\lambda_{j-1}$\\
		\quad\quad\tabitem $\theta_j=\delta\theta_{j-1}$\\
		\quad\quad\tabitem $i=0$\\
		
		\quad\tabitem End \textbf{While}\\

		\hline
		
	\end{tabular}
	
\end{table}

To solve (\ref{MBSL0}), we only need to omit the last term in the gradient value (\ref{SubGrad}) and enforce the constraint   $\textbf{a}'_{N+2:N+L+2}=(-1)_{(L+1) \times 1}$ in each iteration of \textbf{Algorithm 1}. 

For the scenarios in which $K$ is known prior to reconstruction, the modified BIHT algorithm solves (\ref{MBIHT})
\begin{equation}\label{MBIHT}
	\begin{array}{rrclcl}
		\displaystyle \min_{\textbf{a}'} & \multicolumn{3}{l}{\|[Y(\Phi'\textbf{a}')]_-\|_1} \\
		\textrm{s.t.} &&&\\
		& \|\textbf{a}'\|_0 & \leq & K \\
		& \textbf{a}'_{N+2:N+L+2} & = & (-1)_{(L+1) \times 1} \\
	\end{array}
\end{equation}

To solve (\ref{MBIHT}), we propose \textbf{Algorithm 2} which is composed of a Gradient Descent (GD) step followed by projection both onto $\mathcal{C}$ and the $K$-sparse signal space. \textbf{Algorithm 2} provides the stepwise presentation for the BIHT algorithm modified for LC reconstruction.

\section{Simulation results}
\label{sec:sim}
In this section we demonstrate efficient performance of the proposed ZC/LC reconstruction algorithms on random sparse signals generated according to the model presented in \ref{subsec:ZerocrossMod} and provide comparisons with previous works.

\subsection{ZC Reconstruction Performance by 1-Bit CS}\label{subsec:ZCSimul}
\begin{figure}[H]
	\centering
	\includegraphics[scale=.45]{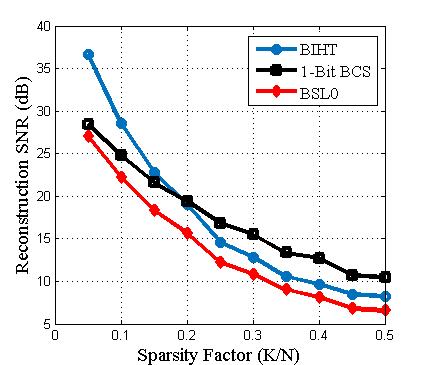}
	\caption{ZC Reconstruction by Different 1-Bit CS Algorithms}
	
	\label{ZCRecon}
\end{figure}

Considering the sparse ZC problem addressed in subsection \ref{subsec:ZerocrossMod}, fig. \ref{ZCRecon} compares the final reconstruction SNR values achieved by the BIHT \cite{1bitJacques}, 1-Bit Bayesian Compressive Sensing (1-Bit BCS) \cite{1bitBCS}, and the proposed BSL0 algorithms. Note that the signal parameters are set as $N=500$, $d=2$ sec, $T=5 \times 10^{-4}$ sec, and $\omega_0=10$ rad/sec and the number of iterations for all algorithms is 50. Also the BSL0 algorithm parameters are set at $\mathbf{(\sigma_0,\lambda_0,\theta_0)}=(0.1, 2.5 \times 10^{-4}, 0.3)$, $\mathbf{(\alpha,\beta,\delta)}=(0.9, 2, 2)$, $\epsilon=0.0005$, $\mu=0.7$, $\sigma_{min}=0.001$. 

\begin{table}[H]
	\centering
	\begin{tabular}{|l|l|}
		\hline
		%		\multicolumn{1}{|c|}{ }\\ 
		\multicolumn{1}{|c|}{\textbf{Algorithm 2} The Binary Iterative Hard Thresholding}\\
		\multicolumn{1}{|c|}{(BIHT) Algorithm Modified for LC Reconstuction}\\ 
		\hline \\	
		\textbf{Inputs}: \\
		\quad\tabitem 	$\Phi'_{M\times(N+1)}$: The sampling matrix\\
		\quad\tabitem $\textbf{y}'_{M\times1}$: The vector of sign measurements\\
		\quad\tabitem \textit{$\epsilon$}: The stopping criteria\\
		\quad\tabitem $IterMax$: The maximum number of iterations\\ 
		\quad\tabitem $\mu$: The step-size to the Gradient Descent (GD)\\   
		
		\textbf{Output}: \\
		\quad\tabitem 
		$\hat{\textbf{a}'}^{(k)}$: The estimated vector of sparse coefficients\\
		\textbf{Algorithm}: \\
		\quad\tabitem Initialization \quad \textit {$\hat{\textbf{a}}^{(1)}=[0]_{(N+1)\times 1}$}, \quad \textit{$k=1$},\\
		\quad \quad \textit{$\hat{\textbf{a}}^{(0)}=[-100]_{(N+1) \times 1}$}\\
		\quad\tabitem \textbf{While} $(||\hat{\textbf{a}'}^{(k)}-\hat{\textbf{a}'}^{(k-1)}||>\epsilon)$ and ($k<IterMax$) \\
		
		\quad \quad  \quad  -~Calculate the gradient vector $\nabla C(\hat{\textbf{a}'}^{(k)})$ \\
		\begin{tabular}{p{8cm}}
			
			%	\tableequation 
			
			{\quad \quad\quad \quad  \quad $\nabla C(\hat{\textbf{a}'}^{(k)})=\frac{1}{2}\Phi'^T(sign(\Phi'\hat{\textbf{a}'}^{(k)})-\textbf{y}')$} 
			
		\end{tabular}\\
		\quad \quad  \quad  -~Perform the gradient descent (GD) step as:\\

		\begin{tabular}{p{8cm}}
			
			%	\tableequation 
			{\quad \quad\quad \quad  \quad $\hat{\textbf{a}'}^{(k+1)}=\hat{\textbf{a}'}^{(k)}-\mu \nabla C(\hat{\textbf{a}'}^{(k)})$}\\ 
			
		\end{tabular}\\
		\quad \quad  \quad   -~Best K-term approximation by thresholding:\\
		
		\begin{tabular}{p{8cm}}
			
			%	\tableequation 
			{\quad \quad\quad \quad  \quad $\hat{\textbf{a}'}^{(k+1)}_{1:N+1} = \eta_{K}(\hat{\textbf{a}'}^{(k+1)}_{1:N+1})$}\\ 
			
		\end{tabular}\\
		
		\quad \quad  \quad   -~Projection onto $\mathcal{C}$ by:\\
		
		\begin{tabular}{p{8cm}}
			
			%	\tableequation 
			{$\quad \quad\quad \quad  \quad \hat{\textbf{a}'}^{(k+1)}_{N+2:N+L+2} = (-1)_{(L+1) \times 1}$}\\ 
			
		\end{tabular}\\
		
		\quad \quad  \quad   -~$k=k+1$ \\
		
		\quad\tabitem End \textbf{While}\\

		\hline
		
	\end{tabular}
\end{table}

Note that although 1-Bit BCS outperforms BIHT and BSL0 for less sparse signals, but its simulation time per iteration was observed to exceed the other two at least by a factor of 10.

\subsection{LC Reconstruction Performance by Modified Signed-CS}\label{subsec:LCSimul}
\begin{figure}[H]
	\centering
	\includegraphics[scale=.4]{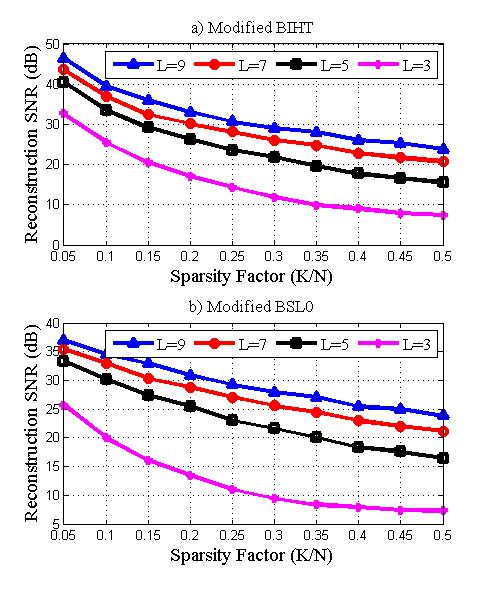}
	\caption{LC Reconstruction SNR for a) Modified BIHT and b) Modified BSL0}
	
	\label{LCRecon}
\end{figure}

Considering the sparse LC problem addressed in subsection \ref{subsec:LevelcrossMod}, fig. \ref{LCRecon} provides the final reconstruction SNR values achieved by the modified BIHT and the modified BSL0 algorithms for different number of reference levels $L$. Note that the signal and algorithm parameters are the same as \ref{subsec:ZCSimul} and the levels are placed uniformly in the dynamic range of the input signal.

\subsection{Comparison with the Literature for Sparse Octave-Band Signals}\label{subsec:OctaveBand}
\begin{figure}[H]
	\centering
	\includegraphics[scale=.4]{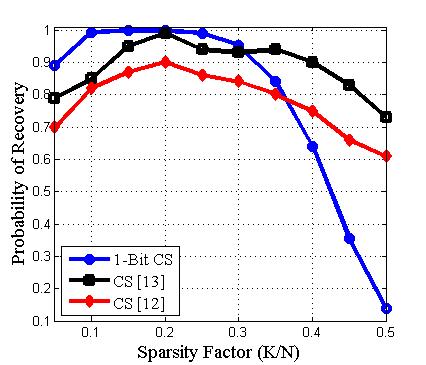}
	\caption{1-Bit vs. Conventional CS for ZC/LC Reconstruction}
	
	\label{Comp}
\end{figure}

As both prior works on sparse ZC/LC reconstruction \cite{Sharma,Boufounos} have considered octave-band signals for simulations, we also report the simulation results for the same scenario for the sake of comparisons. To this end, we limit the harmonics to the interval $n=201,...,400$ and plot the probability of successful recovery by (\ref{onebitprob}) against the sparsity factor in fig. \ref{Comp}. Similar to the literature, the reconstruction SNR values $> 20dB$ are considered as successful recovery in this simulation. Note that this figure compares the performance of the 1-Bit CS approach to ZC reconstruction in this paper with the conventional CS approach taken by \cite{Sharma,Boufounos}. As observed in this figure, migrating to the 1-Bit CS model improves the reconstruction performance for sparser signals while the conventional CS (i.e. \cite{Sharma,Boufounos}) performs better as the sparsity factor increases. 

\section{Conclusion}
\label{sec:con}

In this paper, we have formulated the problem of sparse signal reconstruction from its Level Crossings in terms of 1-bit Compressive Sensing models. We have shown how the LC problem can be addressed by a system of simultaneously constrained signed-CS problems and modified the Binary Iterative Hard Thresholding (BIHT) and Binary Smoothed L0 (BSL0) algorithms to solve this problem. 

\ifCLASSOPTIONcaptionsoff
  \newpage
\fi

\bibliographystyle{IEEEbib}
\bibliography{ref}

\end{document}